\documentclass[aps,prl,reprint,superscriptaddress]{revtex4-1}
\usepackage{graphicx}
\usepackage{color,amsmath}

\usepackage[normalem]{ulem}

\begin{document}


\title{Characterization of spin-orbit interactions of GaAs heavy holes using a quantum point contact}



\author{Fabrizio Nichele}
\email[]{fnichele@phys.ethz.ch}
\homepage[]{www.nanophys.ethz.ch}
\affiliation{Solid State Physics Laboratory, ETH Z\"{u}rich, 8093 Z\"{u}rich, Switzerland}

\author{Stefano Chesi}
\affiliation{CEMS, RIKEN, Wako, Saitama 351-0198, Japan}
\affiliation{Beijing Computational Science Research Center, Beijing 100084, China}

\author{Szymon Hennel}
\affiliation{Solid State Physics Laboratory, ETH Z\"{u}rich, 8093 Z\"{u}rich, Switzerland}

\author{Angela Wittmann}
\affiliation{Solid State Physics Laboratory, ETH Z\"{u}rich, 8093 Z\"{u}rich, Switzerland}

\author{Christian Gerl}
\affiliation{Universit\"{a}t Regensburg, Universit\"{a}tsstrasse 31, 93053 Regensburg, Germany}

\author{Werner Wegscheider}
\affiliation{Solid State Physics Laboratory, ETH Z\"{u}rich, 8093 Z\"{u}rich, Switzerland}

\author{Daniel Loss}
\affiliation{CEMS, RIKEN, Wako, Saitama 351-0198, Japan}
\affiliation{Department of Physics, University of Basel, Basel, Switzerland}

\author{Thomas Ihn}
\affiliation{Solid State Physics Laboratory, ETH Z\"{u}rich, 8093 Z\"{u}rich, Switzerland}

\author{Klaus Ensslin}
\affiliation{Solid State Physics Laboratory, ETH Z\"{u}rich, 8093 Z\"{u}rich, Switzerland}


\date{\today}

\begin{abstract}
We present transport experiments performed in high quality quantum point contacts embedded in a GaAs two-dimensional hole gas. The strong spin-orbit interaction results in peculiar transport phenomena, including the previously observed anisotropic Zeeman splitting and level-dependent effective $g$-factors. Here we find additional effects, namely the crossing and the anti-crossing of spin-split levels depending on subband index and magnetic field direction. Our experimental observations are reconciled in an heavy hole effective spin-orbit Hamiltonian where cubic- and  quadratic-in-momentum terms appear. The spin-orbit components, being of great importance for quantum computing applications, are characterized in terms of magnitude and spin structure. In the light of our results, we explain the level dependent effective $g$-factor in an in-plane field. Through a tilted magnetic field analysis, we show that the QPC out-of-plane $g$-factor saturates around the predicted 7.2 bulk value.
\end{abstract}


\maketitle


Spin-orbit interaction (SOI) is a relativistic effect that couples the motion of an electron to its spin \cite{Ashcroft1976}. For two-dimensional electron gases in the conduction band of III-V materials, SOI originates from bulk inversion asymmetry (Dresselhaus SOI \cite{Dresselhaus1955a}) and structure inversion asymmetry (Rashba SOI \cite{Bychkov1984}) and takes the form $H_{SO}=\beta_D(\sigma_xk_x-\sigma_yk_y)+\alpha_R(\sigma_xk_y-\sigma_yk_x)$, with $\boldsymbol{\sigma}$ the Pauli matrices and $\boldsymbol{k}$ the in-plane wavevector \cite{Ihn2010}. For two-dimensional hole gases (2DHGs) in the valence band of GaAs the situation is very different. Because of the non-zero orbital angular momentum, bulk SOI, and confinement in growth direction, SOI for holes is expected to be more pronounced than for their electronic counterparts, mainly of Rashba type and cubic in $\boldsymbol{k}$ \cite{Winkler2003, Nichele2014}. The relevance of an additional term, quadratic in $\boldsymbol{k}$ and proportional to the in-plane components of the applied magnetic field ${\bf B}$, was recently proposed \cite{bulaev2007electric,chesi2007exchange,Chen2010,Komijani2013a}. Such a term is unique for heavy holes and very useful for exploiting SOI for quantum computing applications \cite{bulaev2007electric}. In this manuscript we show how the cubic and quadratic terms present in the bulk Hamiltonian can be separately addressed in the magnetoconductance of a quantum point contact (QPC) embedded in a 2DHG. Furthermore, our results offer a better understanding of the physics of $p$-type QPCs in terms of level dependent in-plane and out-of-plane $g$-factors ($g_\parallel$ and $g_\perp$ respectively) and allow us to measure the bulk $g_\perp$. The latter is particularly interesting, since the bulk $g$-factor anisotropy of $p$-type GaAs \cite{Kesteren1990,Sapega1992,Marie1999} makes it impossible to directly measure $g_\perp$ with conventional transport techniques \cite{Fang1968}. Theoretical predictions for a [001]-growth 2DHG estimate $g_\parallel=0$ and $g_\perp=7.2$ \cite{Winkler2000a,Winkler2003}. It was argued \cite{Srinivasan2012} that in a QPC, in the limit of high subband index $n$, $g_\perp$ should approach the bulk value. So far, despite the tendency of $g_\perp$ to increase with $n$, this prediction was not experimentally confirmed.

The experiment was performed using a carbon doped GaAs 2DHG grown along the [001] direction. A strong Rashba SOI is expected here due to the asymmetry of the confinement potential. A complete characterization of this 2DHG and its very strong SOI has been reported in Ref. \onlinecite{Nichele2014}.

Three QPCs were defined by electron beam lithography and wet etching and measured by standard low frequency lock-in techniques at a temperature of $100~\rm{mK}$. Two side gates allow independent tuning of the conductance of each QPC. The lithographic width of the QPCs is $240~\rm{nm}$ in one case and $350~\rm{nm}$ for the other two. The QPCs are arranged to form a three-terminal cavity, whose properties will be reported elsewhere. The presented results are independent of the particular structure used. The measurements reported in the following refer to the $350~\rm{nm}$ wide QPC, aligned along the [010] crystallographic direction. The other two devices had similar orientation and showed comparable results. Previous studies on QPCs embedded in [001]-grown 2DHGs did not reveal any dependency of the crystallographic orientation of the QPC \cite{Chen2010,Komijani2013a}. The sample was mounted on a tiltable stage that allowed rotation around one axis. Changing the in-plane magnetic field orientation by $90^\circ$ required warming up the sample and manually changing its bonding configuration. In the color plots that follow we will show the QPC transconductance, that is the numerical derivative of the conductance with respect to the gate voltage axis, in arbitrary units. The color code used is such that a light yellow region indicates a small transconductance (a plateau in the conductance) and a dark blue line indicates high and negative transconductance (the transition region between two plateaus).

\begin{figure}
\includegraphics[width=\columnwidth]{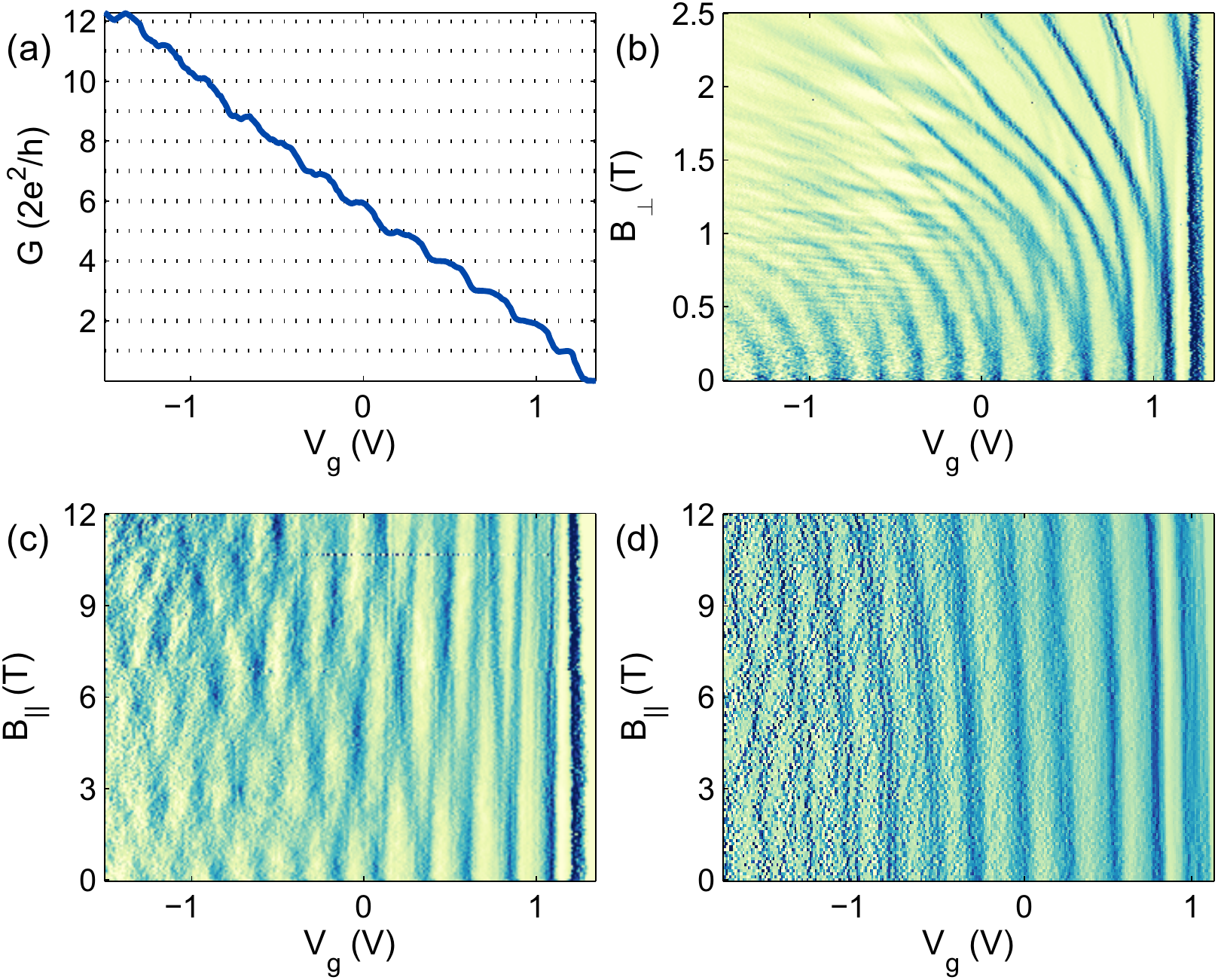}
\caption{(a) QPC linear conductance for $\mathbf{B}=0$ as a function of gate voltage, a gate dependent contact resistance is subtracted from the raw data. (b) Transconductance as a function of gate voltage and $B_\perp$. The arrows point to three examples of anti-crossing (red) and crossing (black). (c) Transconductance as a function of gate voltage and $B_\parallel$, with $B_\parallel$ aligned with the QPC axis. (d) Transconductance as a function of gate voltage and $B_\parallel$, with $B_\parallel$ aligned perpendicular to the QPC axis.}
\label{fig1}
\end{figure}

The QPC zero-field linear conductance is shown in Fig.~\ref{fig1} as a function of the voltage $V_g$ applied to its side gates. In the leakage-free range of the side gates, the QPC shows at least $10$ well developed plateaus. Fig.~\ref{fig1}(b), (c), and (d) show the QPC transconductance for three different magnetic field orientations with respect to the QPC axis. In Fig. \ref{fig1}(b), ${\bf B}$ is applied out-of-plane, in Fig. \ref{fig1}(c) and (d) ${\bf B}$ is applied in-plane with direction parallel and perpendicular to the QPC axis respectively. Similarly to previous work \cite{Zulicke2006,Chen2010,Komijani2013a} a clear Zeeman splitting is present when ${\bf B}$ is out-of-plane or in-plane and oriented along the QPC axis. No evidence of spin splitting up to $12~\rm{T}$ is visible when the field is applied in-plane to the sample and perpendicular to the QPC axis. In an out-of-plane field, in addition to the Zeeman splitting, a bending of the levels towards higher energy (more negative gate voltage) is observed. The latter is due to the formation of magnetoelectric subbands caused by the combined effect of cyclotron energy and confinement potential in the transverse direction \cite{Beenakker1991}.
 
As observed in Refs. \onlinecite{Zulicke2006,Chen2010,Komijani2013a}, the subbands cross in an in-plane magnetic field oriented along the QPC axis [Fig. \ref{fig1}(c)] independently of their quantum numbers. Interestingly, when the field is out-of-plane [Fig. \ref{fig1}(b)], spin-split subbands form a complex pattern where both crossings and anti-crossings appear. The high energy spin-split subbands anti-cross with the low energy spin-split subbands of the neighboring energy level. After the anti-crossing takes place, spin-split levels approach each other and cross. In Fig. \ref{fig1}(b) we mark three examples of anti-crossings (red arrows) and crossings (black arrows). The existence of these anti-crossings for ${\bf B}$ along $z$ (i.e., out-of-plane) suggests a strong influence of an in-plane SOI field, which has a finite matrix elements between the Zeeman eigenstates. On the other hand, the absence of anti-crossings for ${\bf B}$ along $x$ (i.e., along the QPC) is consistent with such SOI being proportional to $\sigma_x$. As we will show below, this interpretation is substantiated through a model describing the interplay of a SOI quadratic in $\bf k$ and the cubic Rashba SOI, which has previously allowed to explain the anomalous spin-polarization observed in hole QPCs through magnetic focusing \cite{chesi2011anomalous}.

We first show the relevance of a quadratic SOI of the type $ H^{(2)}_{SO}= \gamma B_\parallel (p_-^2 \sigma_+ + {\rm h.c.})/2$ \cite{bulaev2007electric, chesi2007exchange} in relation to Fig.~\ref{fig1}(c). To describe transport in the QPC, we assume a harmonic confinement potential along $y$ and a parametric dependence on $x$ of the lateral wavefunction \cite{lesovik1988reflectionless}. The onset of a conductance plateau occurs when $k_x \simeq 0$, at the narrowest point of the QPC constriction, and leads us to consider the following unperturbed Hamiltonian:
\begin{equation}\label{Hx0}
H_{0}=\frac{p_y^2}{2m}+\frac{1}{2}m\Omega^2 y^{2} - \gamma B_\parallel p^2_y \sigma_x -\frac{g_\perp \mu_B}{2} B_\perp \sigma_z,
\end{equation} 
with $p_\pm = p_x \pm i p_y$, $\sigma_\pm=\sigma_x \pm i \sigma_y$,  $\Omega^2=\omega^{2}+\omega_c^{2} $, and  $\omega_c=eB_\perp/m$. Notice that Eq.~(\ref{Hx0}) is restricted to $\bf B$ either  along  $x$ or $z$ (a general $\bf B$ in the  $xz$-plane is discussed later) and is still valid for a more general $ H^{(2)}_{SO}$, which takes into account crystal anisotropy \cite{Komijani2013a}.  

\begin{figure}
\includegraphics[width=\columnwidth]{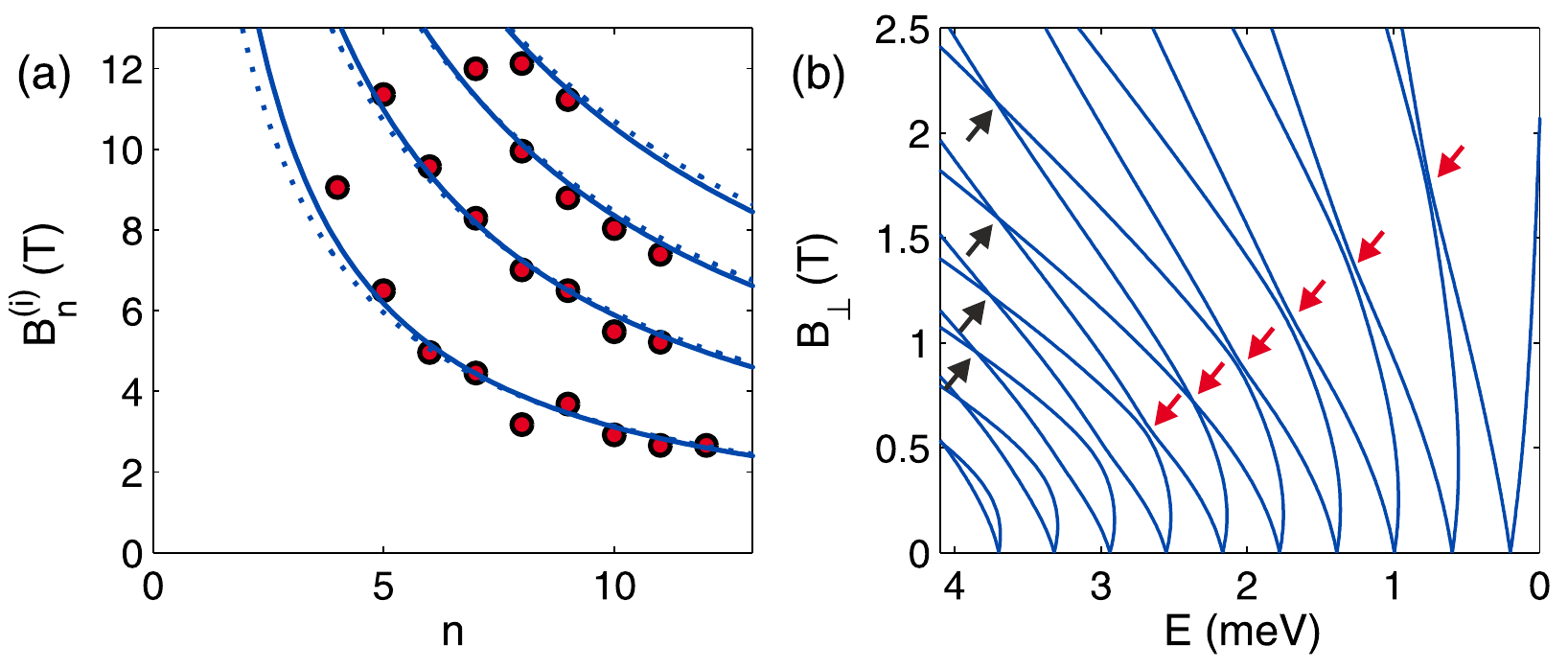}
\caption{(a) Crossing fields $B_n^{(i)}$ extracted from Fig.~\ref{fig1}(c) (dots) together with a fit of Eq.~(\ref{Bni}) assuming harmonic confinement (solid lines) or hard wall confinement (dotted lines). (b) Calculated $k_x=0$, $B_\parallel=0$ eigenvalues of $H_0 +  H_{SO}^{(3)}$. The red arrows indicate anti-crossing points between $n$ and $n+1$ states, the black arrows indicate crossing points between $n$ and $n+2$ states. We used $\hbar \omega=0.4$~meV, $g_\perp = 7.2$, $m=0.3m_0$, and $\hbar^3 \alpha_{2,3} =0.08\times \gamma_{2,3}~{\rm eVnm}^3$ (with $\gamma_{2,3}$ Luttinger parameters), close to the numbers used in Refs.~\onlinecite{chesi2011anomalous,Nichele2014}.}
\label{fig:theory}
\end{figure}

For $B_\perp=0$, the subband energies from Eq.~(\ref{Hx0}) are $E_{n,\pm}=(n-1/2)\hbar\omega \sqrt{1\pm B_\parallel/B_0}$, with $B_0=(2\gamma m)^{-1}$. The values of $B_\parallel$ at which the 1D levels cross are obtained from the condition $E_{n,+}=E_{n+i,-}$:
\begin{equation}\label{Bni}
B_n^{(i)}=B_0\frac{(n+i-1/2)^2-(n-1/2)^2}{(n+i-1/2)^2+(n-1/2)^2},
\end{equation}
and are compared directly to the experimental results. In Fig.~\ref{fig:theory}(a) we perform a fit of Eq.~(\ref{Bni}) (solid line) to the crossing fields obtained from Fig. \ref{fig1}(c) up to fourth order (dots). The theoretical model reproduces the experimental crossings over a wide range of values of $n$ and $B_\parallel$ using a single fitting parameter $B_0=31.2~{\rm T}$. This result is in reasonable agreement with perturbative estimates: using the formulas of Ref.~\onlinecite{Komijani2013a}, for a triangular well with the QPC along [010] and 2D density $n_s=3\times 10^{15}~{\rm m}^{-2}$, we obtain $B_0 \simeq 80~{\rm T}$. The obtained value strongly depends on the detailed form assumed for the confinement \cite{Komijani2013a}.

A valuable feature of Eq.~(\ref{Bni}) is the weak dependence on the specific form of the lateral potential assumed in Eq.~(\ref{Hx0}), which allows one to single out the effect of the quadratic SOI. Equation~(\ref{Bni}) is independent of the QPC harmonic confinement potential $\hbar\omega$. Considering an infinite well is achieved by the replacement $n \to n+1/2$ in Eq.~(\ref{Bni}). The equation in this case is independent of the width $\Delta y$. This introduces a small change in $B_n^{(i)}$ at moderate $n$, and the large-$n$ asymptotic behavior $B_n^{(i)} \simeq B_0 i/n$ is not affected. Therefore, the experimental values of $B_n^{(i)}$ are reproduced with similar accuracy [dotted line in Fig.~\ref{fig:theory}(a)] using $B_0=33~{\rm T}$. In contrast, other quantities are generally rather sensitive to the form of the lateral potential, only approximately known. For example, following the arguments of Ref. \onlinecite{Chen2010, Komijani2013a}, $g_\parallel$ for subband $n$ is obtained as $g_n=(2n-1)\hbar \omega/(\mu_B B_0)$ and $g_n=(n\pi\hbar/\Delta y)^2/(m\mu_B B_0)$ for harmonic and rectangular confinement, respectively. These values have a strong dependence on the confinement parameters $\omega,\Delta y$ as well as on $n$. Independently of the specific confinement chosen, the increasing value of $g_\parallel$ with $n$ finds agreement with the experiment [see the later analysis leading to Fig.~\ref{fig5}(a)].

We now consider $B_\parallel=0$ and the effect of the Rashba SOI. From a third-order perturbative calculation for the two-dimensional subbands, the following form of anisotropic SOI is obtained:
\begin{eqnarray}\label{Hrashba}
 H_{SO}^{(3)} &=&-\left( \alpha_{2}\left\{ p_{y},\left(
p_{x}^{2}-p_{y}^{2}\right) \right\} +2\alpha_{3}\left\{ p_{x},\left\{
p_{x},p_{y}\right\} \right\} \right) \sigma _{x} \nonumber \\
&+&\left( \alpha_{2}\left\{ p_{x},\left( p_{x}^{2}-p_{y}^{2}\right) \right\}
-2\alpha_{3}\left\{ p_{y},\left\{ p_{x},p_{y}\right\} \right\} \right) \sigma _{y},\qquad
\end{eqnarray}
where $\{a,b \}=(ab+ba)/2$ and  $p_{x}=\hbar k_{x}-eB_\perp y$. Taking $\alpha_{2,3}=\alpha$ and $B_\perp=0$, Eq.~(\ref{Hrashba}) recovers the more familiar isotropic expression $i\alpha (p_+^3 \sigma_- - {\rm h.c.})/2 $ \cite{Winkler2000}. 

Taking $k_x \simeq 0$ in Eq.~(\ref{Hrashba}) allows one to explain the anti-crossings observed when $B_\parallel=0$ [red arrows in Fig.~\ref{fig1}(b)] as due to the finite off-diagonal matrix element between Zeeman eigenstates \cite{comment_updown}: $\langle n+1,\uparrow |H_{SO}^{(3)}|n,\downarrow\rangle = i(\Omega + \omega _{c})  \left[ 3\alpha_{2} (\Omega^{2} +\omega _{c}^{2})-2(2\alpha_{2}+\alpha_{3})\Omega \omega _{c}\right] (n\hbar m/2\Omega)^{3/2} $  (for $B_\perp>0$). Interestingly, $\langle n+2,\sigma' |H_{SO}^{(3)}|n,\sigma\rangle=0$ since $H_{SO}^{(3)}$ is odd with respect to $y\to -y$. This feature is in agreement with the higher-order crossings or weak anti-crossings observed in the experiment [black arrows in Fig.~\ref{fig1}(b)]. The full eigenstates of $H_0 + H_{SO}^{(3)}$ for $k_x \simeq 0$ are computed as a function of $B_\perp$ in Fig.~\ref{fig:theory}(b), which is remarkably close to the experiment considering our very simplified modeling of the QPC. Among the other features, our model predicts a non-monotonic behavior of the anti-crossing gaps with $n$, as observed both in Fig.~\ref{fig1}(b) and Fig.~\ref{fig:theory}(b). It arises due to the fact that $\langle n+1,\uparrow |H_{SO}^{(3)}|n,\downarrow\rangle=0$ at a specific value of $B_\perp$. This is only possible if $\alpha_2\neq\alpha_3$, i.e. it is due to the anisotropy in $H_{SO}^{(3)}$. The spin dependence of Eq.~(\ref{Hrashba}) is also in agreement with the absence of anti-crossings in Fig.~\ref{fig1}(c) as, for $B_\perp=0$ and $k_x \simeq 0$, $H_{SO}^{(3)} $ simplifies to $\alpha_2 p_y^3 \sigma_x$ and the spin-orbit perturbation is parallel to $\bf B$. The same is not expected for a general SOI: the Dresselhaus term derived in Ref.~\onlinecite{bulaev2007electric} yields $\beta (p_-p_+p_-\sigma_+ + {\rm h.c.})/2\simeq \beta p_y^3 \sigma_y$, which would induce anti-crossings also when $\bf B$ is parallel to the QPC.

\begin{figure}
\includegraphics[width=\columnwidth]{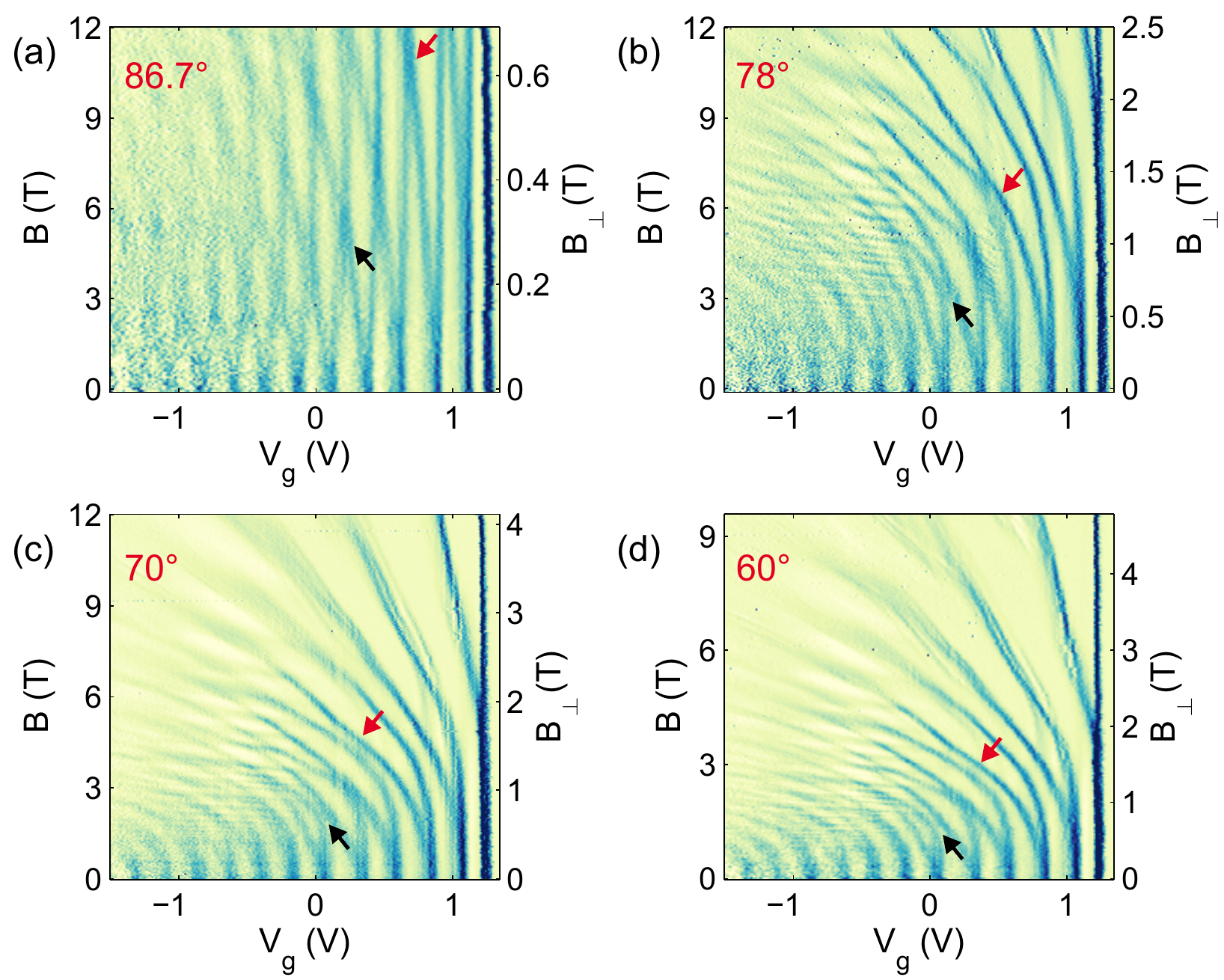}
\caption{QPC transconductance as a function of gate voltage and magnetic field for different tilt angles $\theta$. (a) $\theta=86.7^\circ$. (b) $\theta=78^\circ$. (c) $\theta=70^\circ$. (d) $\theta=60^\circ$. The red and black arrows indicate two features that gradually turn from crossing in (a) to anti-crossing in (d).}
\label{fig3}
\end{figure}

\begin{figure}
\includegraphics[width=\columnwidth]{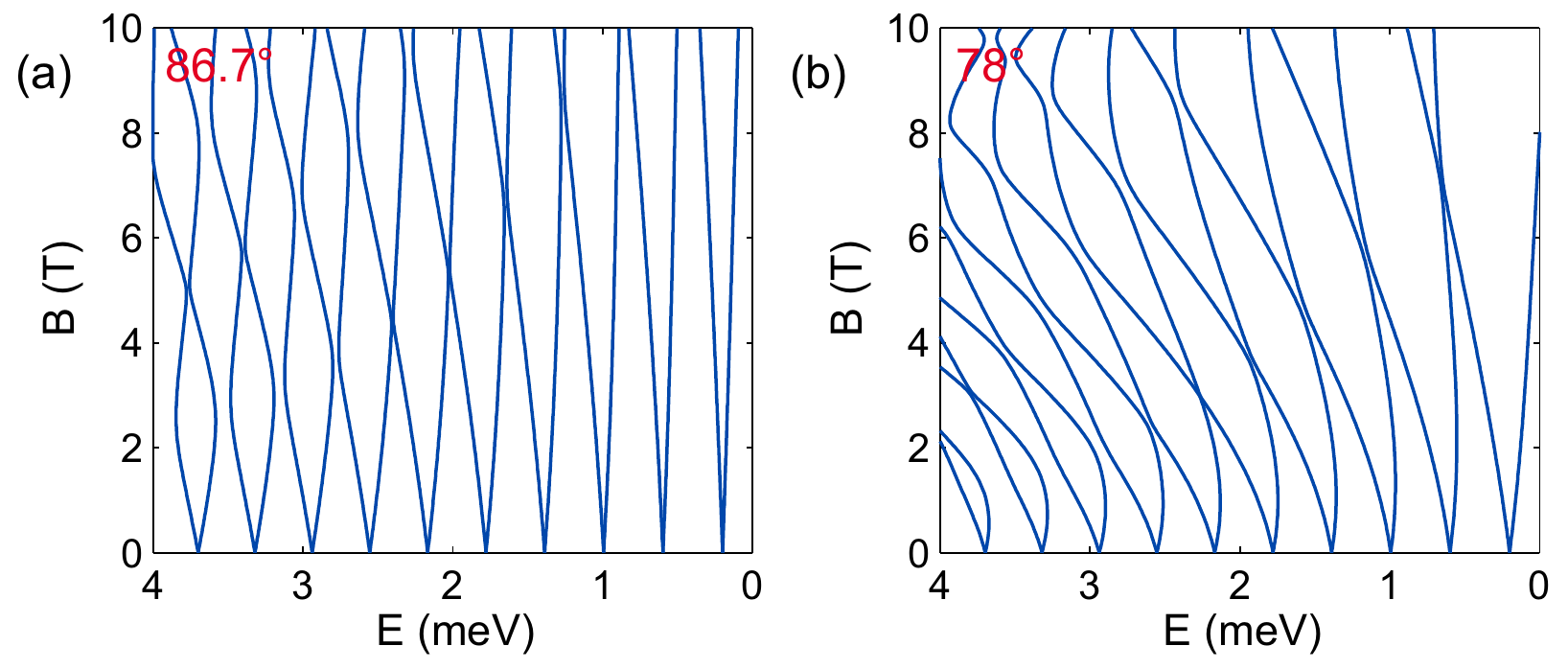}
\caption{Calculated subband energies as in Fig.~\ref{fig:theory}(b) for a finite tilt angle $\theta=86.7^\circ$ (a) and $\theta=78^\circ$ (b). We used an isotropic $H_{SO}^{(2)}$ with a coupling parameter $\gamma$ obtained from $B_0=33$~T.}
\label{angle_theory}
\end{figure}

\begin{figure}
\includegraphics[width=\columnwidth]{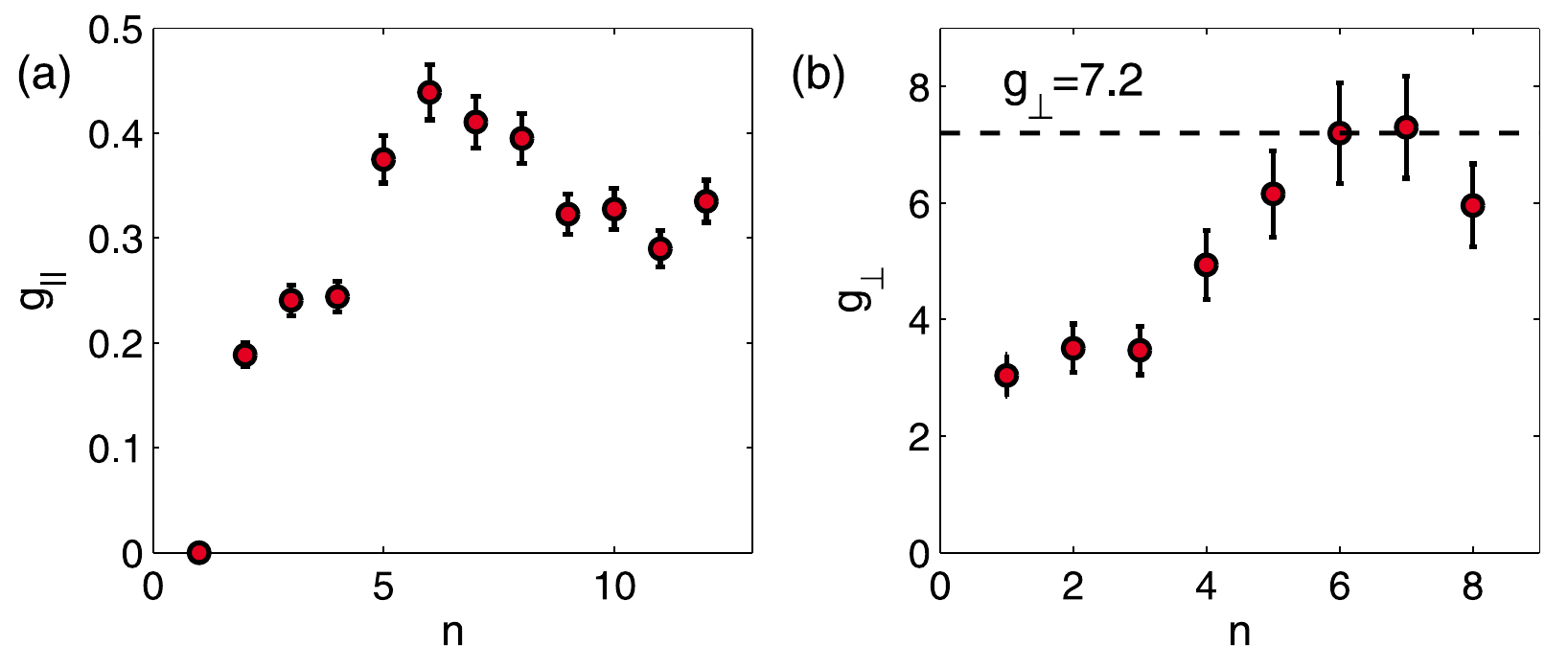}
\caption{(a) Measured $g_\parallel$ as a function of $n$. (b) Measured $g_\perp$ as a function of $n$. The predicted bulk value \cite{Winkler2003} of $g_\perp=7.2$ is indicated by the dashed line.}
\label{fig5}
\end{figure}

After theoretically understanding the effects of a purely in-plane or out-of-plane field, it is interesting to consider the effect of a tilted field. This discussion will lead us to an experimental determination of $g_\perp$. Figure~\ref{fig3} shows four transconductance maps taken for different tilt angles $\theta$ between the 2DHG and the magnetic field, where $\theta$ is indicated in red ($\theta=90^\circ$ indicates a completely in-plane field). The in-plane magnetic field component was kept along the QPC axis. Crossings turn successively into anti-crossings with decreasing tilt angle $\theta$. We mark with a red and a black arrow two of these transitions. For a tilt angle of $86.7^\circ$ [Fig.~\ref{fig3}(a)], the anti-crossings for a subband index $n<8$ tend to become crossings. The high index subbands ($n>8$), that lie at very small in-plane fields, still evidently anti-cross. The features discussed for the $\theta=86.7^\circ$ data are well reproduced in the numerical results of Fig.~\ref{angle_theory}(a). For smaller tilt angles, the general trend observed in Fig.~\ref{fig3} is also found in the simulations as visible in Fig.~\ref{angle_theory}(b). We tested the effect on an in-plane electric field by acquiring transconductance measurements with an asymmetric gate voltage bias. The size of the anti-crossing did not change up to a voltage difference between the two side gates of $1~\rm{V}$, independently of magnetic field tilt angle.

We now turn to the determination of $g_\parallel$ and $g_\perp$. Performing finite bias measurements, we calculated the side gates lever arm $\alpha(V_g)$ and converted the gate voltage axis into an energy axis. In such a way we can directly trace the difference between spin-split levels in Fig. \ref{fig1}(c) as $E_\parallel=g_\parallel\mu_B B_\parallel$ and calculate $g_\parallel$ as a function of $n$. This technique is extensively described in Ref.~\onlinecite{Srinivasan2012} or in the Supplemental Material \cite{Note_supplemental}. In Fig. \ref{fig5}(a) we observe two distinct behaviors: for $n<6$ $g_\parallel$ increases, for $n>6$ it decreases. The initial increasing tendency was observed in numerous experiments \cite{Danneau2006a,Zulicke2006,Chen2010,Srinivasan2012,Komijani2013a} and is explained by the presence of $H_{SO}^{(2)}$. The following tendency reversal naturally originates from the fact that $g_\parallel$ vanishes in the absence of lateral confinement, in a first order approximation. 

The values of $g_\perp$ could not be deduced from Fig.~\ref{fig1}(b) directly due to strong distortions in the linear dependence of the Zeeman splitting introduced by the anti-crossings. Therefore, we measure the level dependent $g_\perp$ using a tilt angle of $86.7^\circ$ [data of Fig.~\ref{fig3}(a)]. This approach is justified since a large in-plane field suppresses the anti-crossings and allows one to extract $g_\perp$ more accurately than at $\theta=0^\circ$ from the linear splitting at low $B_\perp$. Similarly to before, we tracked the levels position as a function of energy and total field $\bf B$, obtaining the total Zeeman splitting $E_Z=\mu_B\sqrt{g_\perp^2 B_\perp^2+g_\parallel^2 B_\parallel^2}$. Using the values from $g_\parallel$ of Fig.~\ref{fig5}(a), we extract $g_\perp$ for $n\leq 8$ with a linear fit of $\sqrt{E_Z^2-\mu_B^2g_\parallel^2B_\parallel^2}$. More details about this procedure are reported in the Supplemental Material \cite{Note_supplemental}. The results are shown in Fig. \ref{fig5}(b) where $g_\perp$ has the tendency to increase with $n$ up to $n=5$ and subsequently saturates. The saturation, concomitant with the decreasing tendency of $g_\parallel$, is compatible with the theoretical expectation of $g_\perp = 7.2$ expected for heavy-holes in a bulk 2DHG \cite{Winkler2003}. The residual presence of anti-crossings for $n\geq 8$ leads to an apparent decrease of Zeeman splitting and a consequent decrease of the extracted $g_\perp$.

The values $g_\perp<7.2$ are consistent with theoretical predictions for narrow 1D wires with a strong heavy-hole light-hole mixing \cite{Zulicke2006}. Similarly reduced values of the bulk $g_\perp$ were measured from the optical spectrum of excitons (e.g., $g_\perp\simeq 2.5$ in Ref.~\onlinecite{Kesteren1990}) and were related as well to light-hole heavy-hole mixing \cite{Sapega1992,Glasberg2001}. In this framework, the level dependence of $g_\perp$ is possibly due to the lower subbands having a stronger lateral confinement, as confirmed by finite bias measurements. A narrow QPC (i.e., with width comparable to the well thickness) cannot be treated through a 2D effective SOI. These limitations of our model might explain the discrepancy between the anomalous behavior of the first plateau and the large Zeeman splitting obtained for $n=1$ in Figs.~\ref{fig:theory}(b) and \ref{angle_theory}. In the regime where the anti-crossings are suppressed and the lateral confinement potential is weak, the experimental $g_\perp$ is comparable to the expected value of $7.2$.

In conclusion, we have investigated the SOI Hamiltonian of a 2DHG using a QPC. With an in-plane field, the level crossings confirm the presence of a quadratic SOI and allow us to measure its direction and strength. The pattern of crossing and anti-crossing observed for an out-of-plane field is compatible with the spin structure of an anisotropic cubic Rashba SOI for heavy holes. Our results give a deeper understanding of spin interactions in 2DHGs, in particular regarding anisotropic effective g-factor and the interplay between different mechanisms of SOI.

\begin{acknowledgments}
The authors wish to thank Roland Winkler and Uli Z\"ulicke for useful discussions and the Swiss National Science Foundation for financial support.
\end{acknowledgments}

\bibliography{Bibliography}

\begin{thebibliography}{27}%
\makeatletter
\providecommand \@ifxundefined [1]{%
 \@ifx{#1\undefined}
}%
\providecommand \@ifnum [1]{%
 \ifnum #1\expandafter \@firstoftwo
 \else \expandafter \@secondoftwo
 \fi
}%
\providecommand \@ifx [1]{%
 \ifx #1\expandafter \@firstoftwo
 \else \expandafter \@secondoftwo
 \fi
}%
\providecommand \natexlab [1]{#1}%
\providecommand \enquote  [1]{``#1''}%
\providecommand \bibnamefont  [1]{#1}%
\providecommand \bibfnamefont [1]{#1}%
\providecommand \citenamefont [1]{#1}%
\providecommand \href@noop [0]{\@secondoftwo}%
\providecommand \href [0]{\begingroup \@sanitize@url \@href}%
\providecommand \@href[1]{\@@startlink{#1}\@@href}%
\providecommand \@@href[1]{\endgroup#1\@@endlink}%
\providecommand \@sanitize@url [0]{\catcode `\\12\catcode `\$12\catcode
  `\&12\catcode `\#12\catcode `\^12\catcode `\_12\catcode `\%12\relax}%
\providecommand \@@startlink[1]{}%
\providecommand \@@endlink[0]{}%
\providecommand \url  [0]{\begingroup\@sanitize@url \@url }%
\providecommand \@url [1]{\endgroup\@href {#1}{\urlprefix }}%
\providecommand \urlprefix  [0]{URL }%
\providecommand \Eprint [0]{\href }%
\providecommand \doibase [0]{http://dx.doi.org/}%
\providecommand \selectlanguage [0]{\@gobble}%
\providecommand \bibinfo  [0]{\@secondoftwo}%
\providecommand \bibfield  [0]{\@secondoftwo}%
\providecommand \translation [1]{[#1]}%
\providecommand \BibitemOpen [0]{}%
\providecommand \bibitemStop [0]{}%
\providecommand \bibitemNoStop [0]{.\EOS\space}%
\providecommand \EOS [0]{\spacefactor3000\relax}%
\providecommand \BibitemShut  [1]{\csname bibitem#1\endcsname}%
\let\auto@bib@innerbib\@empty
\bibitem [{\citenamefont {Ashcroft}\ and\ \citenamefont
  {Mermin}(1976)}]{Ashcroft1976}%
  \BibitemOpen
  \bibfield  {author} {\bibinfo {author} {\bibfnamefont {N.}~\bibnamefont
  {Ashcroft}}\ and\ \bibinfo {author} {\bibfnamefont {N.}~\bibnamefont
  {Mermin}},\ }\href {http://books.google.ch/books?id=FRZRAAAAMAAJ} {\emph
  {\bibinfo {title} {Solid state physics}}}\ (\bibinfo  {publisher} {Saunders
  College},\ \bibinfo {year} {1976})\BibitemShut {NoStop}%
\bibitem [{\citenamefont {Dresselhaus}(1955)}]{Dresselhaus1955a}%
  \BibitemOpen
  \bibfield  {author} {\bibinfo {author} {\bibfnamefont {G.}~\bibnamefont
  {Dresselhaus}},\ }\href {http://link.aps.org/doi/10.1103/PhysRev.100.580}
  {\bibfield  {journal} {\bibinfo  {journal} {Phys. Rev.}\ }\textbf {\bibinfo
  {volume} {100}},\ \bibinfo {pages} {580} (\bibinfo {year}
  {1955})}\BibitemShut {NoStop}%
\bibitem [{\citenamefont {Bychkov}\ and\ \citenamefont
  {Rashba}(1984)}]{Bychkov1984}%
  \BibitemOpen
  \bibfield  {author} {\bibinfo {author} {\bibfnamefont {Y.~A.}\ \bibnamefont
  {Bychkov}}\ and\ \bibinfo {author} {\bibfnamefont {E.~I.}\ \bibnamefont
  {Rashba}},\ }\href {http://stacks.iop.org/0022-3719/17/i=33/a=015} {\bibfield
   {journal} {\bibinfo  {journal} {Journal of Physics C: Solid State Physics}\
  }\textbf {\bibinfo {volume} {17}},\ \bibinfo {pages} {6039} (\bibinfo {year}
  {1984})}\BibitemShut {NoStop}%
\bibitem [{\citenamefont {Ihn}(2010)}]{Ihn2010}%
  \BibitemOpen
  \bibfield  {author} {\bibinfo {author} {\bibfnamefont {T.}~\bibnamefont
  {Ihn}},\ }\href@noop {} {\emph {\bibinfo {title} {Semiconductor
  Nanostructures: Quantum States and Electronic Transport}}}\ (\bibinfo
  {publisher} {Oxford University Press},\ \bibinfo {year} {2010})\BibitemShut
  {NoStop}%
\bibitem [{\citenamefont {Winkler}(2003)}]{Winkler2003}%
  \BibitemOpen
  \bibfield  {author} {\bibinfo {author} {\bibfnamefont {R.}~\bibnamefont
  {Winkler}},\ }\href@noop {} {\emph {\bibinfo {title} {Spin-Orbit Coupling
  Effects in Two-Dimensional Electron and Hole Systems}}},\ \bibinfo {series}
  {Springer Tracts in Modern Physics}, Vol.\ \bibinfo {volume} {191}\ (\bibinfo
   {publisher} {Springer-Verlag, Berlin},\ \bibinfo {year} {2003})\BibitemShut
  {NoStop}%
\bibitem [{\citenamefont {Nichele}\ \emph {et~al.}(2014)\citenamefont
  {Nichele}, \citenamefont {Pal}, \citenamefont {Winkler}, \citenamefont
  {Gerl}, \citenamefont {Wegscheider}, \citenamefont {Ihn},\ and\ \citenamefont
  {Ensslin}}]{Nichele2014}%
  \BibitemOpen
  \bibfield  {author} {\bibinfo {author} {\bibfnamefont {F.}~\bibnamefont
  {Nichele}}, \bibinfo {author} {\bibfnamefont {A.~N.}\ \bibnamefont {Pal}},
  \bibinfo {author} {\bibfnamefont {R.}~\bibnamefont {Winkler}}, \bibinfo
  {author} {\bibfnamefont {C.}~\bibnamefont {Gerl}}, \bibinfo {author}
  {\bibfnamefont {W.}~\bibnamefont {Wegscheider}}, \bibinfo {author}
  {\bibfnamefont {T.}~\bibnamefont {Ihn}}, \ and\ \bibinfo {author}
  {\bibfnamefont {K.}~\bibnamefont {Ensslin}},\ }\href
  {http://link.aps.org/doi/10.1103/PhysRevB.89.081306} {\bibfield  {journal}
  {\bibinfo  {journal} {Phys. Rev. B}\ }\textbf {\bibinfo {volume} {89}},\
  \bibinfo {pages} {081306} (\bibinfo {year} {2014})}\BibitemShut {NoStop}%
\bibitem [{\citenamefont {Bulaev}\ and\ \citenamefont
  {Loss}(2007)}]{bulaev2007electric}%
  \BibitemOpen
  \bibfield  {author} {\bibinfo {author} {\bibfnamefont {D.~V.}\ \bibnamefont
  {Bulaev}}\ and\ \bibinfo {author} {\bibfnamefont {D.}~\bibnamefont {Loss}},\
  }\href@noop {} {\bibfield  {journal} {\bibinfo  {journal} {Phys. Rev. Lett.}\
  }\textbf {\bibinfo {volume} {98}},\ \bibinfo {pages} {097202} (\bibinfo
  {year} {2007})}\BibitemShut {NoStop}%
\bibitem [{\citenamefont {Chesi}\ and\ \citenamefont
  {Giuliani}(2007)}]{chesi2007exchange}%
  \BibitemOpen
  \bibfield  {author} {\bibinfo {author} {\bibfnamefont {S.}~\bibnamefont
  {Chesi}}\ and\ \bibinfo {author} {\bibfnamefont {G.~F.}\ \bibnamefont
  {Giuliani}},\ }\href@noop {} {\bibfield  {journal} {\bibinfo  {journal}
  {Phys. Rev. B}\ }\textbf {\bibinfo {volume} {75}},\ \bibinfo {pages} {155305}
  (\bibinfo {year} {2007})}\BibitemShut {NoStop}%
\bibitem [{\citenamefont {Chen}\ \emph {et~al.}(2010)\citenamefont {Chen},
  \citenamefont {Klochan}, \citenamefont {Micolich}, \citenamefont {Hamilton},
  \citenamefont {Martin}, \citenamefont {Ho}, \citenamefont {Z\"ulicke},
  \citenamefont {Reuter},\ and\ \citenamefont {Wieck}}]{Chen2010}%
  \BibitemOpen
  \bibfield  {author} {\bibinfo {author} {\bibfnamefont {J.~C.~H.}\
  \bibnamefont {Chen}}, \bibinfo {author} {\bibfnamefont {O.}~\bibnamefont
  {Klochan}}, \bibinfo {author} {\bibfnamefont {A.~P.}\ \bibnamefont
  {Micolich}}, \bibinfo {author} {\bibfnamefont {A.~R.}\ \bibnamefont
  {Hamilton}}, \bibinfo {author} {\bibfnamefont {T.~P.}\ \bibnamefont
  {Martin}}, \bibinfo {author} {\bibfnamefont {L.~H.}\ \bibnamefont {Ho}},
  \bibinfo {author} {\bibfnamefont {U.}~\bibnamefont {Z\"ulicke}}, \bibinfo
  {author} {\bibfnamefont {D.}~\bibnamefont {Reuter}}, \ and\ \bibinfo {author}
  {\bibfnamefont {A.~D.}\ \bibnamefont {Wieck}},\ }\href
  {http://stacks.iop.org/1367-2630/12/i=3/a=033043} {\bibfield  {journal}
  {\bibinfo  {journal} {New Journal of Physics}\ }\textbf {\bibinfo {volume}
  {12}},\ \bibinfo {pages} {033043} (\bibinfo {year} {2010})}\BibitemShut
  {NoStop}%
\bibitem [{\citenamefont {Komijani}\ \emph {et~al.}(2013)\citenamefont
  {Komijani}, \citenamefont {Csontos}, \citenamefont {Shorubalko},
  \citenamefont {Z\"ulicke}, \citenamefont {Ihn}, \citenamefont {Ensslin},
  \citenamefont {Reuter},\ and\ \citenamefont {Wieck}}]{Komijani2013a}%
  \BibitemOpen
  \bibfield  {author} {\bibinfo {author} {\bibfnamefont {Y.}~\bibnamefont
  {Komijani}}, \bibinfo {author} {\bibfnamefont {M.}~\bibnamefont {Csontos}},
  \bibinfo {author} {\bibfnamefont {I.}~\bibnamefont {Shorubalko}}, \bibinfo
  {author} {\bibfnamefont {U.}~\bibnamefont {Z\"ulicke}}, \bibinfo {author}
  {\bibfnamefont {T.}~\bibnamefont {Ihn}}, \bibinfo {author} {\bibfnamefont
  {K.}~\bibnamefont {Ensslin}}, \bibinfo {author} {\bibfnamefont
  {D.}~\bibnamefont {Reuter}}, \ and\ \bibinfo {author} {\bibfnamefont {A.~D.}\
  \bibnamefont {Wieck}},\ }\href
  {http://stacks.iop.org/0295-5075/102/i=3/a=37002} {\bibfield  {journal}
  {\bibinfo  {journal} {EPL (Europhysics Letters)}\ }\textbf {\bibinfo {volume}
  {102}},\ \bibinfo {pages} {37002} (\bibinfo {year} {2013})}\BibitemShut
  {NoStop}%
\bibitem [{\citenamefont {van Kesteren}\ \emph {et~al.}(1990)\citenamefont {van
  Kesteren}, \citenamefont {Cosman}, \citenamefont {van~der Poel},\ and\
  \citenamefont {Foxon}}]{Kesteren1990}%
  \BibitemOpen
  \bibfield  {author} {\bibinfo {author} {\bibfnamefont {H.~W.}\ \bibnamefont
  {van Kesteren}}, \bibinfo {author} {\bibfnamefont {E.~C.}\ \bibnamefont
  {Cosman}}, \bibinfo {author} {\bibfnamefont {W.~A. J.~A.}\ \bibnamefont
  {van~der Poel}}, \ and\ \bibinfo {author} {\bibfnamefont {C.~T.}\
  \bibnamefont {Foxon}},\ }\href
  {http://link.aps.org/doi/10.1103/PhysRevB.41.5283} {\bibfield  {journal}
  {\bibinfo  {journal} {Phys. Rev. B}\ }\textbf {\bibinfo {volume} {41}},\
  \bibinfo {pages} {5283} (\bibinfo {year} {1990})}\BibitemShut {NoStop}%
\bibitem [{\citenamefont {Sapega}\ \emph {et~al.}(1992)\citenamefont {Sapega},
  \citenamefont {Cardona}, \citenamefont {Ploog}, \citenamefont {Ivchenko},\
  and\ \citenamefont {Mirlin}}]{Sapega1992}%
  \BibitemOpen
  \bibfield  {author} {\bibinfo {author} {\bibfnamefont {V.~F.}\ \bibnamefont
  {Sapega}}, \bibinfo {author} {\bibfnamefont {M.}~\bibnamefont {Cardona}},
  \bibinfo {author} {\bibfnamefont {K.}~\bibnamefont {Ploog}}, \bibinfo
  {author} {\bibfnamefont {E.~L.}\ \bibnamefont {Ivchenko}}, \ and\ \bibinfo
  {author} {\bibfnamefont {D.~N.}\ \bibnamefont {Mirlin}},\ }\href
  {http://link.aps.org/doi/10.1103/PhysRevB.45.4320} {\bibfield  {journal}
  {\bibinfo  {journal} {Phys. Rev. B}\ }\textbf {\bibinfo {volume} {45}},\
  \bibinfo {pages} {4320} (\bibinfo {year} {1992})}\BibitemShut {NoStop}%
\bibitem [{\citenamefont {Marie}\ \emph {et~al.}(1999)\citenamefont {Marie},
  \citenamefont {Amand}, \citenamefont {Le~Jeune}, \citenamefont {Paillard},
  \citenamefont {Renucci}, \citenamefont {Golub}, \citenamefont {Dymnikov},\
  and\ \citenamefont {Ivchenko}}]{Marie1999}%
  \BibitemOpen
  \bibfield  {author} {\bibinfo {author} {\bibfnamefont {X.}~\bibnamefont
  {Marie}}, \bibinfo {author} {\bibfnamefont {T.}~\bibnamefont {Amand}},
  \bibinfo {author} {\bibfnamefont {P.}~\bibnamefont {Le~Jeune}}, \bibinfo
  {author} {\bibfnamefont {M.}~\bibnamefont {Paillard}}, \bibinfo {author}
  {\bibfnamefont {P.}~\bibnamefont {Renucci}}, \bibinfo {author} {\bibfnamefont
  {L.~E.}\ \bibnamefont {Golub}}, \bibinfo {author} {\bibfnamefont {V.~D.}\
  \bibnamefont {Dymnikov}}, \ and\ \bibinfo {author} {\bibfnamefont {E.~L.}\
  \bibnamefont {Ivchenko}},\ }\href {\doibase 10.1103/PhysRevB.60.5811}
  {\bibfield  {journal} {\bibinfo  {journal} {Phys. Rev. B}\ }\textbf {\bibinfo
  {volume} {60}},\ \bibinfo {pages} {5811} (\bibinfo {year}
  {1999})}\BibitemShut {NoStop}%
\bibitem [{\citenamefont {Fang}\ and\ \citenamefont {Stiles}(1968)}]{Fang1968}%
  \BibitemOpen
  \bibfield  {author} {\bibinfo {author} {\bibfnamefont {F.~F.}\ \bibnamefont
  {Fang}}\ and\ \bibinfo {author} {\bibfnamefont {P.~J.}\ \bibnamefont
  {Stiles}},\ }\href {http://link.aps.org/doi/10.1103/PhysRev.174.823}
  {\bibfield  {journal} {\bibinfo  {journal} {Phys. Rev.}\ }\textbf {\bibinfo
  {volume} {174}},\ \bibinfo {pages} {823} (\bibinfo {year}
  {1968})}\BibitemShut {NoStop}%
\bibitem [{\citenamefont {Winkler}\ \emph {et~al.}(2000)\citenamefont
  {Winkler}, \citenamefont {Papadakis}, \citenamefont {De~Poortere},\ and\
  \citenamefont {Shayegan}}]{Winkler2000a}%
  \BibitemOpen
  \bibfield  {author} {\bibinfo {author} {\bibfnamefont {R.}~\bibnamefont
  {Winkler}}, \bibinfo {author} {\bibfnamefont {S.~J.}\ \bibnamefont
  {Papadakis}}, \bibinfo {author} {\bibfnamefont {E.~P.}\ \bibnamefont
  {De~Poortere}}, \ and\ \bibinfo {author} {\bibfnamefont {M.}~\bibnamefont
  {Shayegan}},\ }\href {http://link.aps.org/doi/10.1103/PhysRevLett.85.4574}
  {\bibfield  {journal} {\bibinfo  {journal} {Phys. Rev. Lett.}\ }\textbf
  {\bibinfo {volume} {85}},\ \bibinfo {pages} {4574} (\bibinfo {year}
  {2000})}\BibitemShut {NoStop}%
\bibitem [{\citenamefont {Srinivasan}\ \emph {et~al.}(2012)\citenamefont
  {Srinivasan}, \citenamefont {Yeoh}, \citenamefont {Klochan}, \citenamefont
  {Martin}, \citenamefont {Chen}, \citenamefont {Micolich}, \citenamefont
  {Hamilton}, \citenamefont {Reuter},\ and\ \citenamefont
  {Wieck}}]{Srinivasan2012}%
  \BibitemOpen
  \bibfield  {author} {\bibinfo {author} {\bibfnamefont {A.}~\bibnamefont
  {Srinivasan}}, \bibinfo {author} {\bibfnamefont {L.~A.}\ \bibnamefont
  {Yeoh}}, \bibinfo {author} {\bibfnamefont {O.}~\bibnamefont {Klochan}},
  \bibinfo {author} {\bibfnamefont {T.~P.}\ \bibnamefont {Martin}}, \bibinfo
  {author} {\bibfnamefont {J.~C.~H.}\ \bibnamefont {Chen}}, \bibinfo {author}
  {\bibfnamefont {A.~P.}\ \bibnamefont {Micolich}}, \bibinfo {author}
  {\bibfnamefont {A.~R.}\ \bibnamefont {Hamilton}}, \bibinfo {author}
  {\bibfnamefont {D.}~\bibnamefont {Reuter}}, \ and\ \bibinfo {author}
  {\bibfnamefont {A.~D.}\ \bibnamefont {Wieck}},\ }\bibfield  {booktitle}
  {\emph {\bibinfo {booktitle} {Nano Letters}},\ }\href {\doibase
  10.1021/nl303596s} {\bibfield  {journal} {\bibinfo  {journal} {Nano Lett.}\
  }\textbf {\bibinfo {volume} {13}},\ \bibinfo {pages} {148} (\bibinfo {year}
  {2012})}\BibitemShut {NoStop}%
\bibitem [{\citenamefont {Z\"ulicke}(2006)}]{Zulicke2006}%
  \BibitemOpen
  \bibfield  {author} {\bibinfo {author} {\bibfnamefont {U.}~\bibnamefont
  {Z\"ulicke}},\ }\href {http://dx.doi.org/10.1002/pssc.200672801} {\bibfield
  {journal} {\bibinfo  {journal} {Phys. Status Solidi (c)}\ }\textbf {\bibinfo
  {volume} {3}},\ \bibinfo {pages} {4354} (\bibinfo {year} {2006})}\BibitemShut
  {NoStop}%
\bibitem [{\citenamefont {Beenakker}\ and\ \citenamefont {van
  Houten}(1991)}]{Beenakker1991}%
  \BibitemOpen
  \bibfield  {author} {\bibinfo {author} {\bibfnamefont {C.}~\bibnamefont
  {Beenakker}}\ and\ \bibinfo {author} {\bibfnamefont {H.}~\bibnamefont {van
  Houten}},\ }in\ \href {\doibase 10.1016/S0081-1947(08)60091-0} {\emph
  {\bibinfo {booktitle} {Semiconductor Heterostructures and Nanostructures}}},\
  Vol.\ \bibinfo {volume} {Volume 44},\ \bibinfo {editor} {edited by\ \bibinfo
  {editor} {\bibfnamefont {H.}~\bibnamefont {Ehrenreich}}\ and\ \bibinfo
  {editor} {\bibfnamefont {D.}~\bibnamefont {Turnbull}}}\ (\bibinfo
  {publisher} {Academic Press},\ \bibinfo {year} {1991})\ pp.\ \bibinfo {pages}
  {1--228}\BibitemShut {NoStop}%
\bibitem [{\citenamefont {Chesi}\ \emph {et~al.}(2011)\citenamefont {Chesi},
  \citenamefont {Giuliani}, \citenamefont {Rokhinson}, \citenamefont
  {Pfeiffer},\ and\ \citenamefont {West}}]{chesi2011anomalous}%
  \BibitemOpen
  \bibfield  {author} {\bibinfo {author} {\bibfnamefont {S.}~\bibnamefont
  {Chesi}}, \bibinfo {author} {\bibfnamefont {G.~F.}\ \bibnamefont {Giuliani}},
  \bibinfo {author} {\bibfnamefont {L.~P.}\ \bibnamefont {Rokhinson}}, \bibinfo
  {author} {\bibfnamefont {L.~N.}\ \bibnamefont {Pfeiffer}}, \ and\ \bibinfo
  {author} {\bibfnamefont {K.~W.}\ \bibnamefont {West}},\ }\href@noop {}
  {\bibfield  {journal} {\bibinfo  {journal} {Phys. Rev. Lett.}\ }\textbf
  {\bibinfo {volume} {106}},\ \bibinfo {pages} {236601} (\bibinfo {year}
  {2011})}\BibitemShut {NoStop}%
\bibitem [{\citenamefont {Glazman}\ \emph {et~al.}(1988)\citenamefont
  {Glazman}, \citenamefont {Lesovik}, \citenamefont {Khmel'nitskii},\ and\
  \citenamefont {Shekhter}}]{lesovik1988reflectionless}%
  \BibitemOpen
  \bibfield  {author} {\bibinfo {author} {\bibfnamefont {L.}~\bibnamefont
  {Glazman}}, \bibinfo {author} {\bibfnamefont {G.}~\bibnamefont {Lesovik}},
  \bibinfo {author} {\bibfnamefont {D.}~\bibnamefont {Khmel'nitskii}}, \ and\
  \bibinfo {author} {\bibfnamefont {R.}~\bibnamefont {Shekhter}},\ }\href@noop
  {} {\bibfield  {journal} {\bibinfo  {journal} {JETP Lett.}\ }\textbf
  {\bibinfo {volume} {48}},\ \bibinfo {pages} {238} (\bibinfo {year}
  {1988})}\BibitemShut {NoStop}%
\bibitem [{\citenamefont {Winkler}(2000)}]{Winkler2000}%
  \BibitemOpen
  \bibfield  {author} {\bibinfo {author} {\bibfnamefont {R.}~\bibnamefont
  {Winkler}},\ }\href {\doibase 10.1103/PhysRevB.62.4245} {\bibfield  {journal}
  {\bibinfo  {journal} {Phys. Rev. B}\ }\textbf {\bibinfo {volume} {62}},\
  \bibinfo {pages} {4245} (\bibinfo {year} {2000})}\BibitemShut {NoStop}%
\bibitem [{com()}]{comment_updown}%
  \BibitemOpen
  \href@noop {} {}\bibinfo {note} {We use $\pm$ and $\uparrow\downarrow$ for
  the eigenstates of $\sigma_{x,z}$, respectively.}\BibitemShut {Stop}%
\bibitem [{Not()}]{Note_supplemental}%
  \BibitemOpen
  \href@noop {} {}\bibinfo {note} {See the Supplemental Material [...], which
  includes Refs.~\cite{Laux1988,Martin2010}.}\BibitemShut {Stop}%
\bibitem [{\citenamefont {Danneau}\ \emph {et~al.}(2006)\citenamefont
  {Danneau}, \citenamefont {Klochan}, \citenamefont {Clarke}, \citenamefont
  {Ho}, \citenamefont {Micolich}, \citenamefont {Simmons}, \citenamefont
  {Hamilton}, \citenamefont {Pepper}, \citenamefont {Ritchie},\ and\
  \citenamefont {Z\"ulicke}}]{Danneau2006a}%
  \BibitemOpen
  \bibfield  {author} {\bibinfo {author} {\bibfnamefont {R.}~\bibnamefont
  {Danneau}}, \bibinfo {author} {\bibfnamefont {O.}~\bibnamefont {Klochan}},
  \bibinfo {author} {\bibfnamefont {W.~R.}\ \bibnamefont {Clarke}}, \bibinfo
  {author} {\bibfnamefont {L.~H.}\ \bibnamefont {Ho}}, \bibinfo {author}
  {\bibfnamefont {A.~P.}\ \bibnamefont {Micolich}}, \bibinfo {author}
  {\bibfnamefont {M.~Y.}\ \bibnamefont {Simmons}}, \bibinfo {author}
  {\bibfnamefont {A.~R.}\ \bibnamefont {Hamilton}}, \bibinfo {author}
  {\bibfnamefont {M.}~\bibnamefont {Pepper}}, \bibinfo {author} {\bibfnamefont
  {D.~A.}\ \bibnamefont {Ritchie}}, \ and\ \bibinfo {author} {\bibfnamefont
  {U.}~\bibnamefont {Z\"ulicke}},\ }\href
  {http://link.aps.org/doi/10.1103/PhysRevLett.97.026403} {\bibfield  {journal}
  {\bibinfo  {journal} {Phys. Rev. Lett.}\ }\textbf {\bibinfo {volume} {97}},\
  \bibinfo {pages} {026403} (\bibinfo {year} {2006})}\BibitemShut {NoStop}%
\bibitem [{\citenamefont {Glasberg}\ \emph {et~al.}(2001)\citenamefont
  {Glasberg}, \citenamefont {Shtrikman},\ and\ \citenamefont
  {Bar-Joseph}}]{Glasberg2001}%
  \BibitemOpen
  \bibfield  {author} {\bibinfo {author} {\bibfnamefont {S.}~\bibnamefont
  {Glasberg}}, \bibinfo {author} {\bibfnamefont {H.}~\bibnamefont {Shtrikman}},
  \ and\ \bibinfo {author} {\bibfnamefont {I.}~\bibnamefont {Bar-Joseph}},\
  }\href {http://link.aps.org/doi/10.1103/PhysRevB.63.201308} {\bibfield
  {journal} {\bibinfo  {journal} {Phys. Rev. B}\ }\textbf {\bibinfo {volume}
  {63}},\ \bibinfo {pages} {201308} (\bibinfo {year} {2001})}\BibitemShut
  {NoStop}%
\bibitem [{\citenamefont {Laux}\ \emph {et~al.}(1988)\citenamefont {Laux},
  \citenamefont {Frank},\ and\ \citenamefont {Stern}}]{Laux1988}%
  \BibitemOpen
  \bibfield  {author} {\bibinfo {author} {\bibfnamefont {S.}~\bibnamefont
  {Laux}}, \bibinfo {author} {\bibfnamefont {D.}~\bibnamefont {Frank}}, \ and\
  \bibinfo {author} {\bibfnamefont {F.}~\bibnamefont {Stern}},\ }\href
  {http://www.sciencedirect.com/science/article/pii/0039602888906711}
  {\bibfield  {journal} {\bibinfo  {journal} {Surface Science}\ }\textbf
  {\bibinfo {volume} {196}},\ \bibinfo {pages} {101} (\bibinfo {year}
  {1988})}\BibitemShut {NoStop}%
\bibitem [{\citenamefont {Martin}\ \emph {et~al.}(2010)\citenamefont {Martin},
  \citenamefont {Szorkovszky}, \citenamefont {Micolich}, \citenamefont
  {Hamilton}, \citenamefont {Marlow}, \citenamefont {Taylor}, \citenamefont
  {Linke},\ and\ \citenamefont {Xu}}]{Martin2010}%
  \BibitemOpen
  \bibfield  {author} {\bibinfo {author} {\bibfnamefont {T.~P.}\ \bibnamefont
  {Martin}}, \bibinfo {author} {\bibfnamefont {A.}~\bibnamefont {Szorkovszky}},
  \bibinfo {author} {\bibfnamefont {A.~P.}\ \bibnamefont {Micolich}}, \bibinfo
  {author} {\bibfnamefont {A.~R.}\ \bibnamefont {Hamilton}}, \bibinfo {author}
  {\bibfnamefont {C.~A.}\ \bibnamefont {Marlow}}, \bibinfo {author}
  {\bibfnamefont {R.~P.}\ \bibnamefont {Taylor}}, \bibinfo {author}
  {\bibfnamefont {H.}~\bibnamefont {Linke}}, \ and\ \bibinfo {author}
  {\bibfnamefont {H.~Q.}\ \bibnamefont {Xu}},\ }\href
  {http://link.aps.org/doi/10.1103/PhysRevB.81.041303} {\bibfield  {journal}
  {\bibinfo  {journal} {Phys. Rev. B}\ }\textbf {\bibinfo {volume} {81}},\
  \bibinfo {pages} {041303} (\bibinfo {year} {2010})}\BibitemShut {NoStop}%
\end{thebibliography}%

\end{document}